\documentclass[preprint,showpacs,preprintnumbers,amsmath,amssymb]{revtex4}

\usepackage{graphicx}
\usepackage{dcolumn}
\usepackage{bm}

\begin{document}


\title{Strain-Control of the magnetic anisotropy \\ in (Ga,Mn)(As,P) ferromagnetic semiconductor layers}
\author{A. Lema\^itre} \email{aristide.lemaitre@lpn.cnrs.fr}
\author{A. Miard}
\author{L. Travers}
\author{O. Mauguin}
\author{L. Largeau}
\affiliation{Laboratoire de Photonique et de Nanostructures, CNRS, route de Nozay, F-91460, Marcoussis, France}

\author{C. Gourdon}
\author{V. Jeudy}
\affiliation{Institut des Nanosciences de Paris, Universit\'{e} Pierre et Marie Curie-Paris 6, CNRS-UMR7588, 
Paris, F-75015 France }

\author{M. Tran}
\author{J.-M. George}
\affiliation{Unit\'{e} Mixte de Physique CNRS-Thales, Route D\'{e}partementale 128, F-91767 Palaiseau C\'{e}dex and Universit\'{e} Paris-Sud, F-91405, 
Orsay, France}

\date{\today}

\begin{abstract}

A small fraction of phosphorus (up to 10~\%) was incorporated in ferromagnetic (Ga,Mn)As epilayers grown on a GaAs substrate. P incorporation allows reducing the epitaxial strain or even change its sign, resulting in strong modifications of the magnetic anisotropy. In particular a reorientation of the easy axis toward the growth direction is observed for high P concentration. It offers an interesting alternative to the metamorphic approach, in particular for magnetization reversal experiments where epitaxial defects stongly affect the domain wall propagation.
\end{abstract}

\maketitle

The strong sensitivity of the magnetic anisotropy to the epitaxial strain is among the remarkable properties of the (Ga,Mn)As ferromagnetic semiconductor. This feature was early pointed out by \citet{Ohno96}. It stems from the carrier-induced nature of the ferromagnetism in this compound, resulting from the exchange interaction between the magnetic moment of the magnetic ions and the itinerant carriers in the valence band.\cite{Dietl00,Dietl01a} The magnetic properties are therefore strongly related to the shape and the filling of this band. In particular, it has been demonstrated both theoretically \cite{Dietl01a,Abolfath01} and experimentally that the magnetic anisotropy is mainly governed by the valence band anisotropy. The epitaxial strain is a main source of anisotropy. Usually,
compressive strain favors an in-plane easy magnetization, as in
the case of a (Ga,Mn)As layer grown on a GaAs substrate, although
this trend may be reversed at low carrier
density.\cite{Sawicki04,Thevenard05,Thevenard07} In counterpart, under
tensile strains the magnetization becomes spontaneously
oriented along the normal to the film. Experimentally the latter situation is usually obtained by depositing a (Ga,Mn)As layer on top of a metamorphic (In,Ga)As buffer grown on a GaAs substrate.\cite{Shen97} This technique was employed to visualize domain walls, \cite{Thevenard06,Dourlat07,Wang07} and their spin-current induced propagation by Kerr imaging. \cite{Yamanouchi04,Yamanouchi06} However the metamorphic growth mode is associated with the formation of dislocations. The misfit dislocations, while remaining at the (In,Ga)As/GaAs interface, induce a cross-hatch pattern that gives rise to very anisotropic domain wall propagation.\cite{Dourlat07} The threading dislocations, emerging in the (Ga,Mn)As layer, also affect the domain wall propagation by creating pinning centers resulting in filamentary 360$^{\circ}$ domain wall structures.\cite{Thevenard06} It is therefore desirable to fabricate magnetic layers with tensile strain, using a pseudomorphic approach. This would extend the range of possibilities to fabricate advanced magnetic structures such as defect-free magnetic tracks for high-speed domain wall propagation, spin-polarized opto-electrical devices\cite{OhnoY99} with spin polarisation along the growth direction at zero magnetic field, multi-layer structures combining different magnetic anisotropies... In this letter we report on the effect of incorporating a small amount of phosphorus, less than 10~\%, in substitution of the As atoms to modifiy the epitaxial strain within the magnetic (Ga,Mn)(As,P) layers grown by molecular beam epitaxy. Upon sufficient P incorporation we observe a flipping of the magnetization easy axis from in-plane to out-of-plane. Moreover we are also able to visualize the domain wall propagation by Kerr microscopy. We should note that a similar effect has already been recently reported in samples fabricated by combined Mn$^+$-P$^+$ ion implantation in GaAs substrates.\cite{Stone07}

\begin{figure}[h]
\includegraphics[width=8cm]{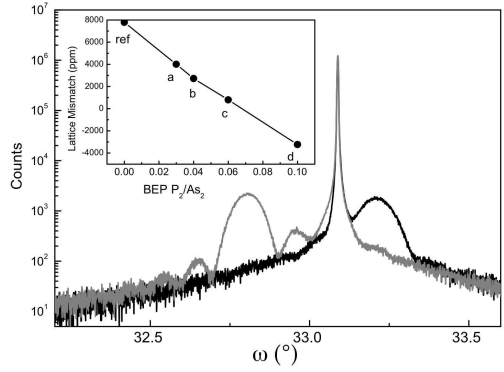}
\caption{\label{XRAY} High resolution X-ray $\omega -2\Theta$ curves around the $(0 0 4)$ reflection for samples \emph{ref} (grey) and \emph{d} (black). Inset: (Ga,Mn)(As,P) lattice mismatch evolution with the BEP $P_2/As_2$ ratio.}
\end{figure}

A series of four (Ga,Mn)(As,P) samples, labelled \emph{a, b, c, d} were grown with increasing P concentrations along with a reference (Ga,Mn)As sample (\emph{ref}). They consist of a 50~nm thick magnetic layer deposited on a GaAs (001) substrate. They were prepared in a very similar conditions as the ones usually employed for (Ga,Mn)As layers. In particular the growth temperature was set around 250$^{\circ}$C for the magnetic layer while the GaAs buffer was grown at 600$^{\circ}$C. As$_2$ was supplied by a valved cracker cell, while P$_2$ was evaporated from a GaP effusion cell. The As flux was chosen so as to be close to the stoichiometric conditions with respect to the Ga and Mn flux for the reference sample. The As$_2$ flux was kept similar for all the samples. The P$_2$ flux was controlled by the cell temperature. The $P_2/As_2$ ratio reported in the following is the ratio of the beam equivalent pressures (BEP) measured by a nude ion gauge prior to the magnetic layer growth. In this series $P_2/As_2$ ranged from 0 to $0.1$. For $P_2/As_2=0.1$ corresponding to sample \emph{d}, the P concentration was estimated from an additional Ga(As,P) reference layer grown under the same conditions but without Mn. The lattice mismatch  $(\Delta a/a_\text{sub})_\text{Ga(As,P)}=(a_\text{perp}-a_\text{sub})/a_\text{sub}=-5790$~ppm was determined by high resolution X-ray diffraction ($a_\text{perp}$ is the lattice parameter along the growth axis of the strained layer, while $a_\text{sub}$ is the GaAs lattice parameter). If the Vegard's law for Ga(As,P) compounds remains valid even at low growth temperature the P concentration would be around 8.8~\%. The high resolution X-ray rocking curves around the $(0 0 4)$ reflection are shown on Fig.~\ref{XRAY} for samples \emph{ref} and \emph{d}. The sharp peak around 33.1$^{\circ}$ corresponds to the diffraction by the GaAs substrate. For the sample \emph{ref} a wider peak is also observed along with some small Pendellosung oscillations. Both features correspond to the diffraction by the (Ga,Mn)As layer of finite thickness. The (Ga,Mn)As diffraction peak is located on the lower angle side, compared to the GaAs peak. $a_\text{perp}$ is therefore larger than $a_\text{sub}$; the (Ga,Mn)As layer is under compressive strain on GaAs. In this sample $(\Delta a/a_\text{sub})_\text{\emph{ref}} = 7790$~ppm is high, indicating that a rather large amount of Mn atoms has been inserted in the matrix (close to 9.5~\%). On the opposite, the (Ga,Mn)(As,P) layer exhibits a larger peak and on the higher angle side, which corresponds to a layer under tensile strain with $(\Delta a/a_\text{sub})_d = -3230$~ppm.  $\Delta a/a_\text{sub}$  was measured for all the samples and is reported in the Fig.~\ref{XRAY} inset as a function of $P_2/As_2$. The strain varies linearly with $P_2/As_2$. As $P_2/As_2$ increases the lattice mismatch decreases, goes close to zero and even changes its sign. Thus, it is possible, by directly controlling the P flux, to tune the amount and the sign of the strain inside the magnetic layer and even to cancel it. At this point, one should notice that the lattice mismatch measured for the highest $P_2/As_2$ layer is much larger, even with opposite sign, than the one expected from the sum of the lattice mismatches of the (Ga,Mn)As and Ga(As,P) layers grown with respectively similar Mn and P concentrations $(\Delta a/a_\text{sub})_\text{\emph{ref}}+(\Delta a/a_\text{sub})_\text{Ga(As,P)}=+2000$~ppm. The origin of this decrepancy is still unclear. In (Ga,Mn)As the lattice mismatch is mostly induced by the incorporation of a non negligeable part of Mn interstitials.\cite{Masek04} These results would therefore suggest that P favors a rather large reduction of the interstitial concentration. This point obviously calls for further investigations.\cite{Glas04}

\begin{figure}[h]
\includegraphics[width=9cm]{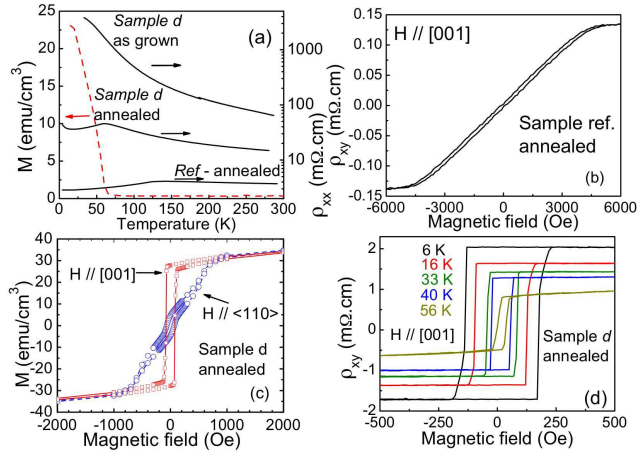}
\caption{\label{Fig2}(a) Temperature dependence of the remanent magnetization (dash line, left axis) and the longitudinal resistivity (solid lines, right axis) for annealed sample \emph{d}. For comparison, the resisitivity is also plotted for as-grown sample \emph{d} and annealed sample \emph{ref}. (b) Transverse resistivity \emph{vs} magnetic field, applied along the growth axis for the reference sample at 4~K. (c) Magnetization vs magnetic field, applied out-of-plane along $[001]$ (squares) or in-plane along $\langle110\rangle$ (circles) at 10~K. (d) Transverse resistivity \emph{vs} magnetic field, applied $[001]$, for sample~\emph{d} at different temperatures.}
\end{figure}

As mentionned above, a change of the strain sign would result in the change of the magnetic anisotropy. This is indeed what is observed. In the following, we compare the magnetic and transport properties of samples \emph{ref} and \emph{d}. In order to measure the longitudinal and transverse resistivities, the samples were processed into Hall bars. In thin (Ga,Mn)As layers low temperature annealing is highly beneficial to the magnetic and transport properties, as it results in the decrease of detrimental Mn interstitial. This is also the case in (Ga,Mn)(As,P). The samples were annealed for 1 hour at 250$^{\circ}$C under N$_{2}$ atmosphere. As seen from the temperature dependence of the longitudinal resistivity (Fig.~\ref{Fig2}(a)), the as-grown sample exhibits a insulating character. In opposite the annealed sample resistivity only shows a moderate increase when lowering the temperature, with a characteristic peak at the ferromagnetic transition around 60~K. The Curie temperature $T_C\simeq 60$~K was determined from the remanent magnetization temperature dependance (Fig~\ref{Fig2}(a)) using a superconducting quantum interference device (SQUID). It is very close to the value determined from the longitudinal resistivity. No magnetic trace of MnAs clusters were detected at higher temperature. The annealed reference sample \emph{ref} has a lower resistivity and the peak occurs at a higher temperature, around 143~K (Fig.~\ref{Fig2}(a)). P incorporation results therefore in the degradation of the conductivity and the Curie temperature. The latter has already been pointed out by \citet{Stone07} and attributed to stronger hole localization. Nevertheless, the samples in this series contain a rather large amount of Mn, requiring a corresponding large amount of P to reverse the strain sign. Lower Mn and P concentrations may result in higher $T_C$, while remaining in tensile strain.

We now turn to the magnetic anisotropy, investigated using the magnetic field dependence of the magnetization, measured by SQUID, and the Hall effect. The latter, in (Ga,Mn)As layers, is dominated by the anomalous term proportional to the perpendicular-to-the-film component of the magnetization. Fig.~\ref{Fig2}(b) shows, over a rather large magnetic field excursion ($\pm6$~kOe), the tranverse resistivity of the reference sample, without P, at 4~K. The magnetic field was applied along the $[001]$ direction. A smooth evolution is observed, corresponding the magnetization being orientated progressively away from its (in-plane) easy axis. Such a magnetization curve is observed up to $T_C$. In this case, the $[001]$ direction is not an easy axis for the magnetization, as expected for (Ga,Mn)As layers in compressive strain on GaAs. Sample \emph{d}, containing the largest P concentration, with a high tensile strain, exhibits a different behavior as seen in Figs.~\ref{Fig2}(c,d). Fig.~\ref{Fig2}(c) shows the magnetization vs magnetic field measured for both orientations, out-of-plane, along $[001]$ and in-plane along $\langle110\rangle$ at 10~K. A well defined square hysteresis loop is now observed when the field is along the $[001]$ growth direction, while the magnetization increases progressively for the $\langle110\rangle$ direction. Therefore, a reorientation of the easy axis, from in-plane to out-of-plane, has occured. It has been induced by the change of the strain sign upon P incorporation. A similar behavior is observed for the transverse resistivity, as seen in Fig~\ref{Fig2}(d), for all temperatures up to $T_C$, namely a square hysteresis loop when the magnetic field is applied along the growth axis. Let us note that the anisotropy field, required to orientate the magnetization along the hard axis, is lower in (Ga,Mn)(As,P) ($\sim760$~Oe) than in the reference sample ($\sim4400$~Oe). The strain absolute value is indeed smaller, the hole density probably lower (lower $T_C$) and therefore result in a lower anisotropy field.\cite{Dietl01a}

\begin{figure}[h]
\includegraphics[width=8cm]{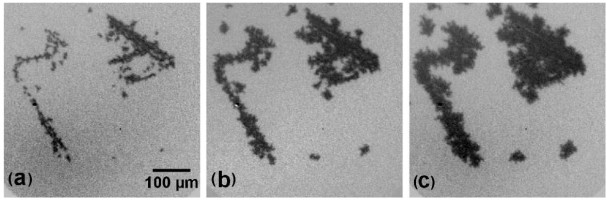}
\caption{\label{Kerr} Successive snapshots of the magnetization reversal in sample \emph{d} at 5 K, observed by Kerr microscopy, after having applied a series of 2 (a), 6 (b), 10 (c) field pulses of amplitude -100 Oe and duration 1s. }
\end{figure}

Last, some preliminary Kerr microscopy measurements of magnetic domain reversal were performed to illustrate the potentiality of this technique. The experimental details were already published.\cite{Dourlat07, Gourdon07} The images shown on Fig.~\ref{Kerr} are successive snapshots of the magnetization reversal taken after applying a series of -100~Oe pulses, with 1 second duration. A magnetic contrast is clearly visible, black and grey regions corresponding to domains with opposite perpendicular-to-the-plane magnetization. Upon magnetic pulses, the magnetic configuration changes, black domains growing at the expense of domains with magnetization opposite to the magnetic field direction.

In conclusion, the incorporation of P in (Ga,Mn)As layers allows the fine tuning of the epitaxial strain. This results in turn in modification of the magnetic anisotropy, in particular a switching of the magnetic axis from in-plane to out-of-plane for sufficiently large P concentration. This effect is directly related to the carrier-mediated nature of the ferromagnetic phase.

This work has been supported in parts by the ANR PNano project MOMES.

\end{document}